\documentclass[fleqn,10pt]{wlscirep}
\usepackage{amsmath}
\usepackage{nccmath}
\title{The role of the strain induced population imbalance in Valley polarization of graphene: Berry curvature perspective}

\author[1,2]{Tohid Farajollahpour}
\author[1,2,*]{Arash Phirouznia}
\affil[1]{Department of Physics, Azarbaijan Shahid Madani
	University, 53714-161, Tabriz, Iran}
\affil[2]{Condensed Matter
	Computational Research Lab. Azarbaijan Shahid Madani University,
	53714-161, Tabriz, Iran}

\affil[*]{phirouznia@azaruniv.edu}

\begin{abstract}
Real magnetic and lattice deformation gauge fields have been investigated in 
honeycomb lattice of graphene. The coexistence of these two gauges will induce 
a gap difference between two valley points ($K$ and $K'$) of system. This gap 
difference allows us to study the possible topological valley Hall current and 
valley polarization in the graphene sheet. In the absence of magnetic field, 
the strain alone could not generate a valley polarization when the Fermi energy 
coincides exactly with the Dirac points. Since in this case there is not any imbalance 
between the population of the valley points. In other words each of these gauges alone 
could not induce any topological valley-polarized current in the system at zero Fermi energy. 
Meanwhile at non-zero Fermi energies population  imbalance can be generated as a 
result of the external strain even at zero magnetic field. In the context of Berry curvature within the linear 
response regime the valley polarization (both magnetic free polarization, 
$\Pi_0$, and field dependent response function, $\chi_\alpha$) in different 
values of gauge fields of lattice deformation has been obtained.
\end{abstract}

\begin{document}

\flushbottom
\maketitle
%
%
\thispagestyle{empty}
Valleytronics, the valley version of spintronics is based on quantum valley number which carries the information by valley degree of freedom \cite{rycerz2007valley}. The family of two dimensional materials with hexagonal lattice structures are potentially a good candidate to manipulate valley polarization \cite{rycerz2007valley,isberg,zhu2012field,goswami2007,yang2013spin,salfi,xiao2007valley}. Recently the first observation of valley Hall effect in graphene family with inversion symmetry breaking has been reported \cite{sui2015gate,shimazaki2015}. Various proposals have been 
presented for generation of the valley Hall conductivity \cite{schaibley2016}. Massive Dirac electrons in graphene has been suggested for valleytronics applications \cite{gappedFilter1,gappedFilter2}.   
In the literatures, most of the proposals for valley-filtering realization are combinations of  
strain induced gauge fields and electromagnetic fields 
\cite{low2010strain,fujita2010valley,zhai2011valley,zhai2010magnetic,chaves}. A valley filter has been proposed  regarding the fact that the effective magnetic field (that results from combination of the strain and external magnetic field) is large in one of the Dirac cones and can be zero in the other \cite{chaves}. In the presence of Rashba spin orbit coupling and magnetic barrier 
in strained graphene, it is possible to produce valley and spin 
polarized currents \cite{wu2016full}. Jiang and $et~al$ proposed 
a scheme for generation of valley current in strained graphene in 
which the strain could be described by cyclic adiabatic deformations and there is 
a chemical potential in suspended region\cite{jiang2013}. 
Wang and $et~al$ have used the method of adiabatic quantum 
pumping in a three barrier structure with strained graphene 
and a ferromagnetic layer to generate pure valley current\cite{wang2014}. 

The heart of the field of straintronic in two dimensional materials is 
to manipulate the valley polarization by applying various deformations. The possibility of strain induced pseudomagnetic field in graphene family
up to 300 T is a remarkable result and getting attention to strain engineering in graphene-like systems
\cite{levy2010strain,klimov2012electromechanical,zhu2015programmable}.
Further challenges for generating pseudomagnetic fields in graphene and
other two dimensional materials not restricted to locally strained
graphene nanobubbles. There is recently another proposal
by Zhu and coworkers in which a pseudomagnetic field around 200 T has been obtained by
an uniaxial stretched graphene~\cite{zhu2015programmable}. Arias and
coworkers considered the possibility of gauge field generation from an elastic deformation in graphene by a
quantum field theory approach. They found a relation between the
pseudomagnetic field and Riemann curvature \cite{arias2015gauge}. Vaezi and
coworkers by means of time dependent strain and consequently a time dependent
gauge field showed that the charge current could be generated due to the time
dependent elastic deformations in graphene \cite{vaezi2013topological}. By specific 
configurations of external potential \cite{da2015valley} or strain \cite{yan2013strain} , it is possible to access valley 
polarized current in bilayer graphene . In addition, the valley 
dependent Hall transport in the presence of electric field was proposed 
before in Ref. \cite{xiao2007valley}. In this letter we propose a device based on 
magnetic field and nonuniform strain which could be applied as depicted in Fig.~\ref{gap}.(a),\ref{gap}.(b),\ref{gap}.(c). We study the valley polarization of the sample in the context of Berry curvature. By considering the gap difference between two valley points ($K$ and $K'$) 
the possible topological valley Hall current and valley polarization are studied.

\section*{Methodology}\label{valley}
\begin{figure}
	\centering
	\includegraphics[width=1.00\linewidth]{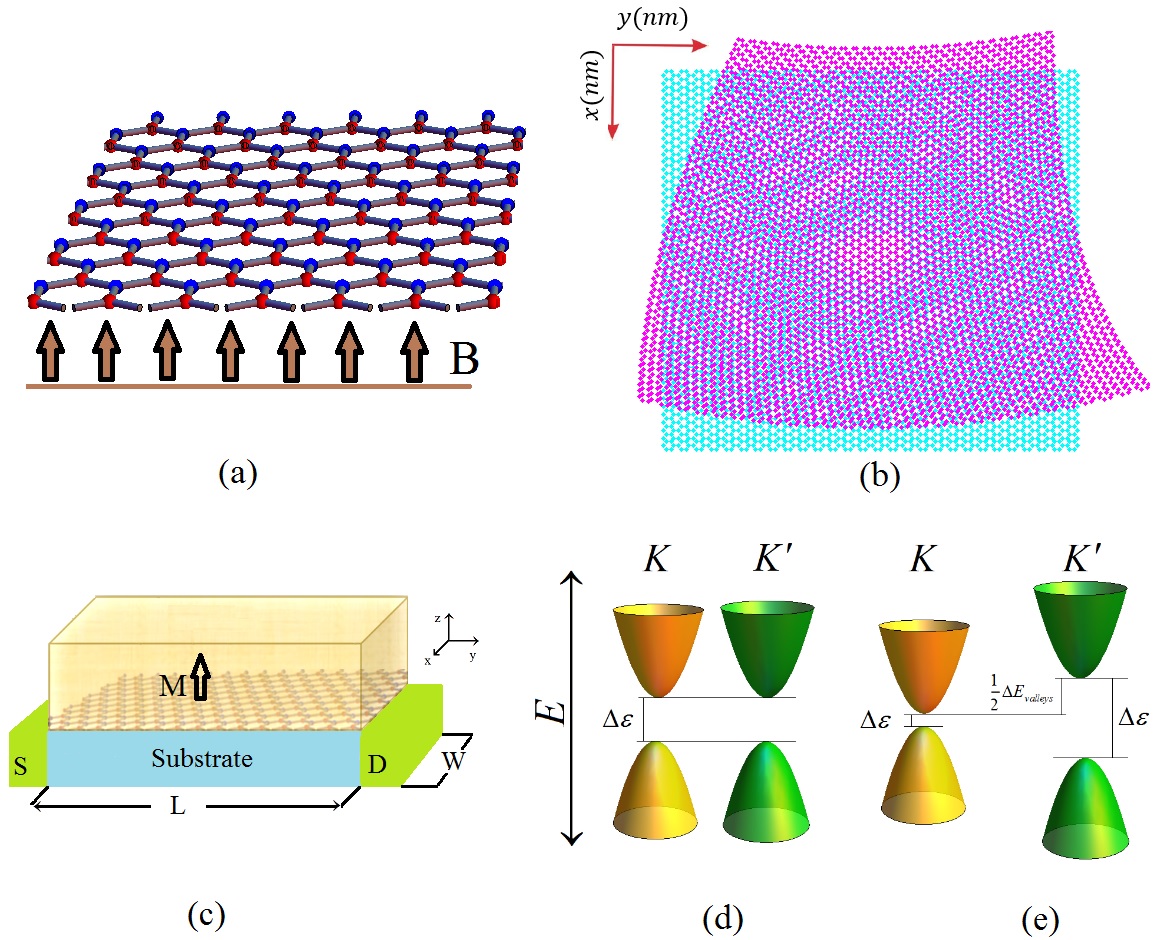}
	\caption{(Color online)  (a) Gapped graphene sample in the presence of magnetic field, (b) gapped graphene sample under proposed deformation, (c) 
		a schematic proposal for the valley Hall current and valley polarization in graphene. A ferromagnetic metal with $z$ direction of magnetization has been placed on top. The central 
		region is the strained graphene (the special form of strain is imposed by substrate. 
		The S and D are the source and drain respectively. W and L are the width and length 
		of strained graphene (d), (e) The valley gap difference (VGP) 
		created in a gapped graphene 
		in the presence of electromagnetic field and strain. 
		(d) When the system has a spectral gap ($\Delta \mathcal{E}$)  
		around each of valleys, $K$ and $K'$. In the absence of magnetic field in the sample, the strain alone could not generate any 
		gap difference between these valley points in $k_f=0$, where $k_f$ is the radius of the Fermi circle. As  illustrated the 
		value of spectral gaps in two valleys are 
		equal. (e) Applying a magnetic field in strained graphene, 
		induces a gap difference $\Delta E_{valleys}$ between the valleys. 
		For typical parameters of $A_{S_x}=2A_{S_y} = -0.008~eV$ , 
		$A_{M_x} = A_{M_y} = 0.0024~eV$, $m =5~ meV$, $\Delta \mathcal{E}_K = 0.0206~eV$, 
		$\Delta \mathcal{E}_{K'} = 0.0616~eV$  the gap difference 
		is $\Delta E_{valleys} = 0.041~eV$. For more information about two gauges of strain ${A}_{S}$ and magnetic field $A_M$ see the supplementary information. } 
	\label{gap}
\end{figure}
Both strain and 
magnetic field could be realized as experimental tuning  parameters \cite{neek2012a,neek2012b,neek2012c,exp1}.   
Although the magnetic field breaks the time reversal symmetry, however, 
it cannot lift the valley degeneracy. Simultaneous presence of magnetic field and 
strain in gapped graphene leads to a gap difference between two valleys.   
Despite the various proposals where have been based on combinations 
of strain, magnetic and electric fields 
\cite{low2010strain,fujita2010valley,zhai2011valley,zhai2010magnetic,wu2016full} in the current study, the proposed sample is based on a magnetic field and a nonuniform strain. The
valley-resolved Hamiltonian in k-space reads
\begin{ceqn} 
\begin{eqnarray}
H^{\eta}_{gr}(k)=\hat{\Psi}_{\eta,k}^{\dagger}v_f(\eta \tau_x k_x+\tau_y k_y)\hat{\Psi}_{\eta,k}
\end{eqnarray}
\end{ceqn}
where $v_f$ is Fermi velocity, the valley index $\eta$ is $+ (-)$ for $K (K')$ and the matrices of $\tau$ act on sub-lattice indices (A and B).The strain and real magnetic field contributions in k-space given by
\begin{ceqn} 
\begin{eqnarray}
H^{\eta}_{S}(k)=\hat{\Psi}_{\eta,k}^{\dagger}v_f(\tau_x {A}_{S_x}+{\eta}\tau_y {A}_{S_y})\hat{\Psi}_{\eta,k}
\end{eqnarray}
\end{ceqn}
\begin{ceqn} 
\begin{eqnarray}
\label{Hmk}
H^{\eta}_{M}(k)=\sum_q\hat{\Psi}_{\eta,k+q}^{\dagger} v_f( {\eta}\tau_x {A}_{Mx}+\tau_y {A}_{My})\hat{\Psi}_{\eta,k}
= \sum_{q}\mathcal{H}^{\eta k}_M(q)\nonumber,
\end{eqnarray}
\end{ceqn}
where ${A}_{S}$ is the fictious gauge field due to the strain and $A_M$ is the real magnetic gauge fields (see supplementary information). we have defined
\begin{ceqn} 
\begin{eqnarray}
\mathcal{H}^{\eta k}_M(q)=\hat{\Psi}_{\eta,k+q}^{\dagger} v_f({\eta} \tau_x {A}_{Mx}+\tau_y {A}_{My})\hat{\Psi}_{\eta,k},
\end{eqnarray}
\end{ceqn}
in which $\mathcal{H}^{\eta k}_M(q)$ stands for the magnetic field induced momentum transfer of $q$. Finally the k-space mass term is expressed as
\begin{ceqn} 
\begin{eqnarray}
H^{\eta}_{mass}(k)=\hat{\Psi}_{\eta,k}^{\dagger} (m \tau_z)\hat{\Psi}_{\eta,k}.
\end{eqnarray}
\end{ceqn}
From theoretical point of view a gap opening in the presence of non-uniform strain is \cite{manes2013}, $H_{mass} \propto B^{ps} = \left(\partial_y(u_{xx}-u_{yy}) + 2\partial_xu_{xy}\right)\tau_z$. The energy gap as function of pseudomagnetic field is the Zeeman
coupling of pseudospin to the associated pseudomagnetic field \cite{manes2013}, $E_{Zeeman} = \frac{3}{8} V' a^2 B^{ps}$ 
where $a$ is the lattice constant, and $V' = 6 (eV/\AA)$ \cite{ferone}.

It should be noted that the nature of the magnetic and strain dependent gauges are quite different.
The magnetic gauge field depends on electronic positions, however, the strain induced gauge is a function of the atomic positions ($u_\alpha$).
Unlike the other contributing terms in the Hamiltonian since the magnetic vector potential depends on electrons position in real space, therefore, the Hamiltonian of magnetic field, as indicated in Eq. (\ref{Hmk}), contains momentum transfer contributions and cannot be represented in block diagonal form of independent $k$ subspaces. However, within the first order perturbation approach, in which the momentum transferring terms cannot contribute, it can be shown that the previous results could be considered reliable at the qualitative level \cite{wu2016full}. Unlike the first order correction, momentum transferring terms contribute in the higher orders of the perturbation and  generally used block diagonalization approach cannot give rise to an exact answer. Then the non-perturbative part of the  strained graphene Hamiltonian is
\begin{ceqn} 
\begin{eqnarray}\label{H0}
H^{(0)}=\sum_{\eta}(H^\eta_{gr} + H^\eta_{mass}+H^\eta_S).
\end{eqnarray}
\end{ceqn}
The matrix representation of $H^{(0)}$ 
in the following k-space basis  
\begin{ceqn} 
\begin{eqnarray}\hat{\Psi }_k=\left( \begin{matrix}
{{{\hat{\Psi }}}_{kA}^+}  \\
{{{\hat{\Psi }}}_{kB}^+}\\
{{{\hat{\Psi }}}_{kA}^-}  \\
{{{\hat{\Psi }}}_{kB}^-}
\end{matrix} \right), \label{basis}
\end{eqnarray}
\end{ceqn}
can be written as
\begin{ceqn} 
\begin{eqnarray}\label{H2}
H^{(0)}=\left[ \begin{matrix}

m & P^- & 0 & 0  \\

P^+ & -m & 0 & 0  \\

0 & 0 & -m & Q^-  \\

0 & 0 & Q^+ & m  \\

\end{matrix} \right].
\end{eqnarray}
\end{ceqn}
where $P^{\pm}=v_f(k^{\pm}+{A}^{\pm}_s)$,  $Q^{\pm}=-v_f(k^{\pm}-{A}^{\pm}_s)$, $k^{\pm}=k_x\pm ik_y$ and ${A}_{S}^{\pm}={A}_{S_x}\pm i{A}_{S_y} $. First order perturbation accounts for the zero momentum transfer contribution of the magnetic field. Therefore it can be described as
\begin{ceqn} 
\begin{eqnarray}
H^{(1)} = \sum_{\eta}\mathcal{H}^{\eta k}_M(0).
\end{eqnarray}
\end{ceqn}
This can be represented in a single k-block of the given basis  in the following form 
\begin{ceqn} 
\begin{eqnarray}\label{HM}	H^{(1)}=\left[ \begin{matrix}
0 & ev_fA^-_M & 0 & 0  \\

ev_fA^+_M & 0 & 0 & 0  \\

0 & 0 & 0 & -ev_fA^-_M  \\

0 & 0 & -ev_fA^+_M & 0  \\

\end{matrix} \right]
\end{eqnarray}
\end{ceqn}
where $A_{M}^{\pm}=A_{M_x}\pm iA_{M_y} $. The full Hamiltonian up to the first order perturbation could be written as,
\begin{ceqn} 
\begin{eqnarray}
H=H^{(0)} + H^{(1)}.
\end{eqnarray}
\end{ceqn}
It should be noted that as mentioned before $H^{(1)}$ corresponds to the first order corrections and the higher orders of perturbative corrections can be achieved by taking into account the non-zero momentum transfer contributions. The gap difference between two valleys, $K$ and $K'$, can be obtained within the perturbation theory (see supplementary information). It can be shown that in strained graphene this valley gap difference depends on both magnetic field and applied strain. Within the first order perturbation
valley gap difference, which has been defined as $\Delta E_\textit{VGD}= E^{gap}_K-E^{gap}_{K'}$, 
can  be easily found to be
\begin{ceqn} 
\begin{equation}
\Delta E_\textit{VGD}=\frac{e v_f B L^2}{2} \sum_{\eta}\left(\mathcal{M}_{+\eta}-\mathcal{M}_{-\eta} \right)\neq 0
\end{equation}
\end{ceqn}
where $\mathcal{M}_{\tau\eta}=\mathcal{C}_{\tau\eta}+\mathcal{C}^*_{\tau\eta}$ and $\mathcal{C}_{\tau \eta}=\alpha_{\tau \eta} \beta^*_{\tau \eta}$. The external magnetic field is denoted by $B$ and $L$ stands for the size of the system. The $\alpha$ and $\beta$ are the components of the unperturbed Dirac Hamiltonian eigenvectors, $\psi_{\eta}^{(0)}$ and (see supplementary information for more details),
\begin{ceqn} 
\begin{align}
\label{E0}
\mathcal{E}_{ \eta,\textbf{k}}^{(0) \pm}= \pm \sqrt{m^2+ v_f^2 (k^++\eta A_s^+)(k^-+\eta A_s^-)}
\end{align}
\end{ceqn}
 \begin{figure}
 	\centering
 	\includegraphics[width=0.500\linewidth]{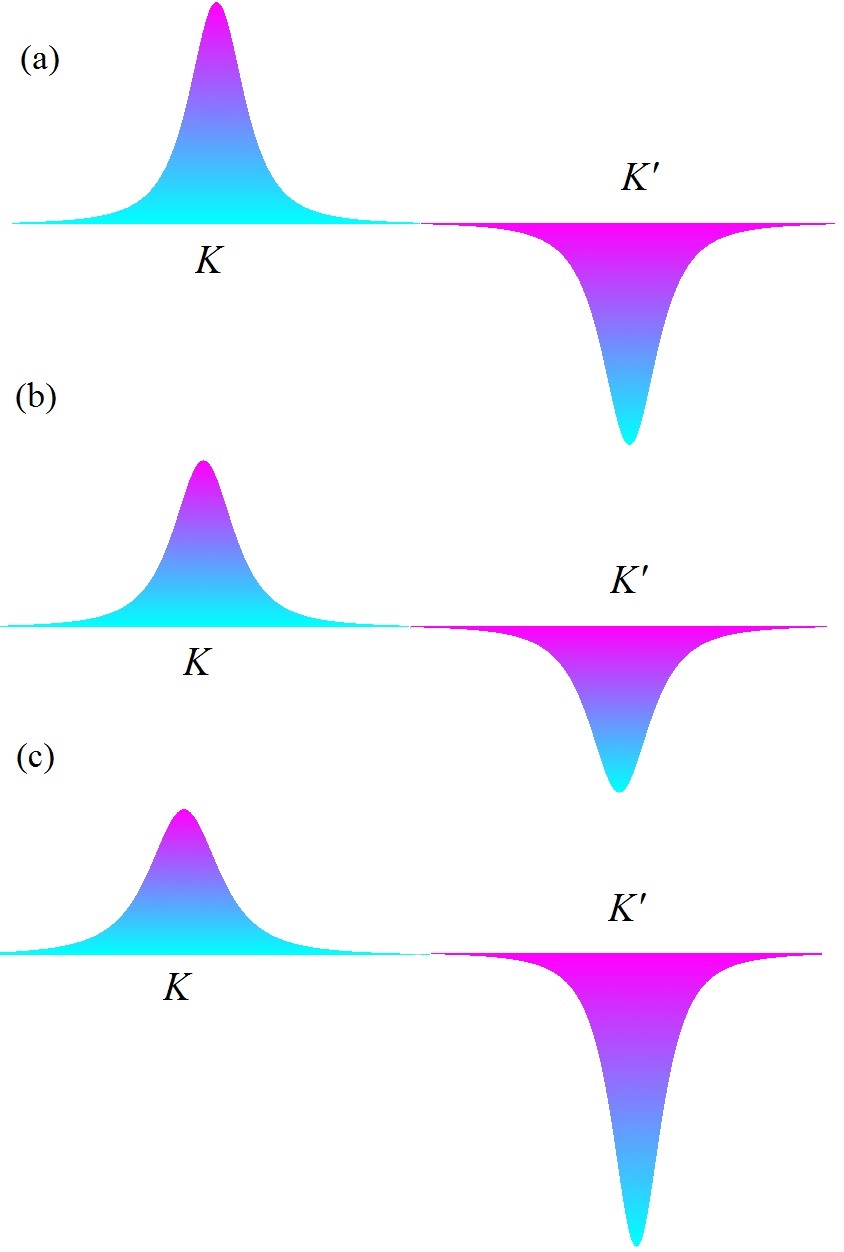}
 	\caption{(Color online) Berry curvature of two valley points $K$ and $K'$. (a) In the absence of strain and magnetic field the Berry curvature of two $K$ and $K'$ are $\Omega^K(k,0,0) = -\Omega^{K'}(k,0,0)$, (b) in the presence of magnetic field $\Omega^K(k,A_m,0) =-\Omega^{K'}(k,A_m,0)$ , (c) in the presence of magnetic field and strain $\Omega^K(k,A_m,A_s) \neq -\Omega^{K'}(k,A_m,A_s)$. 
 		  }
 		  \label{BerryFig}
 \end{figure}
As illustrated in Figs.~\ref{gap}.d and \ref{gap}.e the valley gap difference has been appeared as a result of the magnetic field and strain. It is very important to note that Eq. (\ref{E0}) indicates that for a given $\textbf{k}\neq 0$ state we have $\mathcal{E}_{ K,\textbf{k}}^{(0) \pm}\neq\mathcal{E}_{ K',\textbf{k}}^{(0) \pm}$ even at zero magnetic field. This means that there is a strain induced population imbalance between different valleys at $B = 0$. It has also been realized that the strain 
moves the Dirac points slightly. Meanwhile, this movement does not change the direct band gap of graphene. A valley Hall current can be generated by the population imbalance as a consequence of the valley gap difference.   
We have proposed a practical method for measuring the valley Hall current in Fig.~\ref{gap}.c that could be available using the present experimental techniques. 
By means of this setup, we demonstrated that the valley 
polarized current can be achieved in strained graphene in the presence of 
magnetic field which can be expressed in terms of Berry curvature. 
The topological response to 
an external gauge field could be obtained by integrating 
of Berry curvature of filled bands over the momentum space \cite{bernevig2013topological},
\begin{ceqn} 
\begin{eqnarray}\label{sigmaG}
\sigma_{xy}=\sum_{\eta}\frac{e^2}{2\pi h}\int \int dk_x dk_y~ {{\Omega}_{k_xk_y}^\eta}
\end{eqnarray}
\end{ceqn}
The Berry curvature $\Omega_{k_xk_y}$ of strained graphene in the presence of 
magnetic field is (see supplementary information),
\begin{ceqn} 
\begin{align}\label{valcurv}
\Omega^\eta_{k_xk_y}=\frac{\eta m v^2_f}{\left( v_f^2 \pi_x^2 + v_f^2\pi_y^2 + m^2\right)^{\frac{3}{2}}}
\end{align}
\end{ceqn}
where $\pi_x=k_x+eA_{M_x}+\eta {A}_{S_x}$, $\pi_y=k_y+eA_{M_y}+\eta {A}_{S_y}$. Dependence of Berry curvature to gauge fields is shown in figures \ref{BerryFig}.a,\ref{BerryFig}.b and \ref{BerryFig}.c. Fig.\ref{BerryFig}.a shows the Berry curvature of the system in the absence of any gauge fields. 
\section*{Results and discussion}
\begin{figure}
	\centering
	\includegraphics[width=0.60\linewidth]{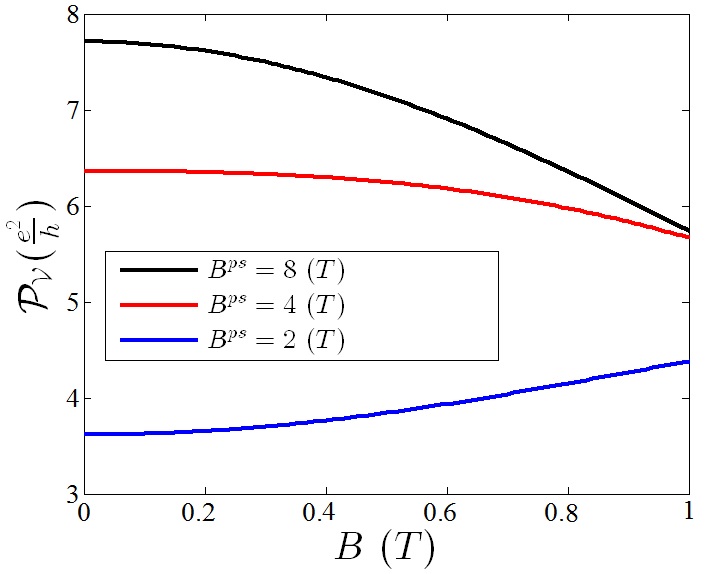}
	
	\caption{(Color online) Valley polarization function in terms of magnetic field. The nonzero $\mathcal{P_V}$ at zero magnetic field ($B=0$) arises as a result of the population imbalance induced by nonzero strain when the radius of each of the Fermi circles is not zero. The $B^{ps}$ is the value of pseudo magnetic field which generate by introduced non-uniform strain (the details are presented in supplementary information)}
	\label{polarization}
\end{figure}
\begin{figure}
	\centering
	\includegraphics[width=.60\linewidth]{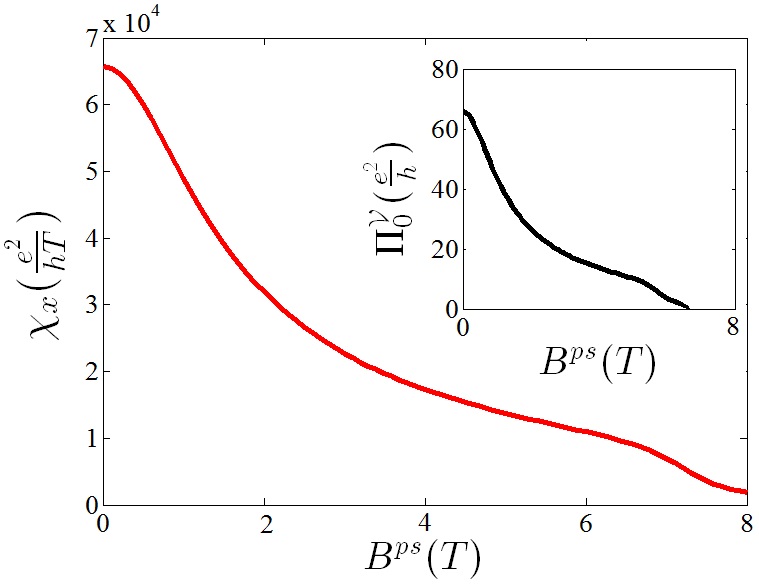}
	
	\caption{(Color online) Linear response function of valley polarization in terms of pseudo magnetic field. (inset) Magnetic free polarization function ($\Pi_0^{\mathcal{V}}$) in terms of pseudo magnetic field.   }
	\label{Response}
\end{figure}
In the absence of the strain, there is not any gap difference 
between two valleys ($E^{gap}_K= E^{gap}_{K'}$). 
Using the relations which were presented in Eqs. \ref{sigmaG} and \ref{valcurv} it can be inferred that the charge conductivity of inequivalent valleys satisfy
$\sigma^{KC}_{xy}=-\sigma^{K'C}_{xy}$. Where $\sigma^{\eta C}_{xy}$ is the contribution of the $\eta$-valley in the charge Hall  conductivity. Meanwhile, the valley resolved current
with a given valley index, $\eta$, can be defined by $j^{\eta}_\mathcal{V}=\langle \eta v \rangle$ where $v$ is the velocity operator.  Accordingly, 
$j^\eta_\mathcal{V}=\eta j_C$ where $j_C$ is the charge current. The valley polarization in the presence of the external magnetic field is then given by $\mathcal{P}_\mathcal{V}^M = {\sigma_\mathcal{V}^{K}-\sigma_\mathcal{V}^{K'}}$ where the valley conductivity $\sigma_\mathcal{V}^{\eta} = \eta \sigma_{C}^{\eta} $. 

As long as the population balance between the two valley points is maintained, valley current cannot be induced in the sample. Valley population imbalance can be achieved when the chemical potential of inequivalent Dirac points are not the same where  
for the strained graphene and in the
presence of magnetic field this is really the case. It should be mentioned that valley population imbalance has been employed for generation of valley current in the 
presence of magnetic field in strained graphene as described in Ref.~\cite{wu2016full}. Furthermore, realization of the valley Hall current strongly depends on population imbalance at different valleys \cite{xiao2007valley}. Accordingly, the population imbalance that comes from the coexistence of real and pseudo-magnetic fields \cite{wu2016full} automatically guaranties the generation of valley polarization in strained valley Hall systems. In the present work it was shown that when the Fermi energy has not exactly been located at Dirac points the strain itself could results in population imbalance and magnetic free valley polarization.\

Since the Berry curvatures of different valleys have opposite sign then the Berry curvature dependent quantities such as Hall conductivity vanishes identically once the population balance is established. This is the key point 
for generation of topological valley current in the system.  Using the Eq. \ref{valcurv} it can be shown that, in the absence of the 
strain, Berry curvature of different Dirac points in the presence of an external magnetic field are opposite. In the other words 
the total charge conductivity vanishes exactly i.e. $\sigma^{KC}_{xy}+\sigma^{K'C}_{xy} = 0$ \cite{vaezi2013topological} (or equivalently valley Hall polarization vanishes $\sigma_\mathcal{V}^{K}-\sigma_\mathcal{V}^{K'}=0$). Unlike the strained sample in this case the position of the Fermi energy cannot result in non-zero charge conductivity.   
In strained honeycomb structure, the presence of magnetic field induces a gap difference between two valley points ($\Delta E_\textit{VGD}\neq 0$). 
This gap difference  leads population imbalance in two inequivalent Dirac points. 
Meanwhile unlike the previous case, the Berry curvature of each valleys doesn't contribute oppositely   and therefore
$\sigma^{CK} \neq -\sigma^{CK'}$ ,
\begin{ceqn} 
\begin{eqnarray}
\mathcal{P}_\mathcal{V} = {\sigma_\mathcal{V}^{K}-\sigma_\mathcal{V}^{K'}}\neq 0.
\end{eqnarray}
\end{ceqn}\

In the absence of the magnetic field the total conductivity vanishes just when the Fermi energy is exactly located at the Dirac points 
(which has been indicated by $k_x=0$ and $k_y=0$ in Eq. \ref{valcurv} which indicates the contribution of the Dirac points). 
In Fig.~\ref{polarization} the valley 
polarization of strained graphene has been depicted in terms of magnetic field. By increasing the value of magnetic fields, 
functionality of valley polarization for each value of 
$B^{ps}$ changes in different way. It should be noted that the zero magnetic polarization which has been shown in this figure comes from the strain induced population imbalance in nonzero Fermi circle as discussed before. We have chosen $E_F\simeq 0.1$eV that corresponds to n-type doping with conduction electron density about $n\simeq 7.3\times 10^{11}cm^{-2}$ and Fermi wave vector $k_f\simeq 0.0152 \AA^{-1}$. 

The linear valley polarization  
response function, $\chi_{\alpha\beta}$, could be obtained 
in the presence of magnetic field. In this case the valley polarization can be given by $\mathcal{P}_{\mathcal{V} \alpha}=\chi_{\alpha\beta} \cdot A_{m_\beta} $. 
At the limit of low magnetic fields by expanding each valley resolved Berry curvatures around $A_M=0$. With the assumption of $A_S > A_M $ 
we have $\mathcal{P}_{\mathcal{V}} = \Pi_0^{\mathcal{V}} + \chi_\alpha~ \delta A_{M_\alpha}$ where 
$\Pi_0^{\mathcal{V}}$ is the magnetic free polarization 
and $\chi_\alpha^{\mathcal{V}}$ is linear response function of 
the strained system. After expansion,
\begin{ceqn}   
\begin{eqnarray}\label{expanded}
 \Omega_k (A_m,A_s)  = \Omega_k (0,A_s)+\frac{\partial \Omega_k(A_m,A_s)}
{\partial A_{m_x}}|_{_{A_{m_x}=0}}\delta A_{m_x} +\frac{\partial \Omega_k(A_m,A_s)}{\partial A_{m_y}}|_{_{A_{m_y}=0}}\delta A_{m_y}
\end{eqnarray} 
\end{ceqn}
substituting  Eq.~\ref{expanded} in Eq.~\ref{sigmaG} and integrating, 
the magnetic free valley polarization and linear response function of the system can be obtained as
\begin{ceqn} 
\begin{eqnarray}
\Pi_0^{\mathcal{V}}&=&\frac{e^2}{2\pi h}\sum_{\eta}\int \int dk_x dk_y~ {{\Omega}_{k_xk_y}^\eta}(0,A_s)\\
\chi_\alpha^{\mathcal{V}}&=&\frac{e^2}{2\pi h}\sum_{\eta}\int \int dk_x dk_y~\frac{\partial \Omega^\eta_{k_xk_y}(A_m,A_s)}{\partial A_{m_\alpha}}|_{_{A_{m_\alpha}=0}}.\nonumber
\end{eqnarray} 
\end{ceqn} 
Within a numerical calculation it can be shown that, depending on the value of the Fermi energy, the  magnetic free valley polarization $\Pi_0^{\mathcal{V}}$ cannot go all the way to zero.  Fig.~\ref{Response} (inset) shows the 
magnetic free polarization in terms of pseudo magnetic field. The linear response function of the magnetic field induced polarization, $\chi_{\alpha}$, in terms of the $B^{ps}$ has also been illustrated in 
Fig.~\ref{Response}. It should be mentioned that because of 
the symmetric form of magnetic gauge field within the Dirac cone approximation, in which the band anisotropy of the sample has been ignored, the behavior of $\chi_y$ is 
expected to be identical with $\chi_x$. 
Strong magnetic field modifies the electronic spectrum 
and forms landau levels. In this case, the magnetic field itself induces valley density polarization in the presence of non-vanishing intrinsic orbital magnetic moment \cite{cai}. 
Since the landau levels of two-dimensional structures are localized states, therefore this polarized density could not be extracted form the system. However the external magnetic field here in the present case is weak enough that cannot  change the spectrum significantly. In addition we have shown that the population imbalance can be induced by deformation gauge even at zero magnetic field.

Berry curvature dependent quantities can arise provided that the inversion symmetry is broken. Meanwhile since the contribution of each valley has been canceled out by its counterpart there is no valley polarization in the graphene based Hall current system. 
Meanwhile as discussed in the current study strain could induce population imbalance between different Dirac points when the radius of the Fermi circle is not zero. Since the valley degeneracy has been lifted by the strain for $\textbf{k}\neq0$ even at $B=0$. This is due to the fact that in the presence of the strain we have  $\mathcal{E}_{ K,\textbf{k}}^{(0) \pm}\neq\mathcal{E}_{ K',\textbf{k}}^{(0) \pm}$. 
This means that when the energy of the Fermi level does not exactly meet the Dirac points there is a population imbalance between two nonequivalent Dirac cones. Therefore in this case valley polarized Berry curvature dependent quantities, such as valley polarized Hall conductivity, could be realized as a result of this population imbalance.

\bibliography{ref}

\section*{Author contributions statement}
All authors contributed equally to this work.

\section*{Additional information}
Competing financial interests: The authors declare no competing financial interests.

\end{document}